\documentstyle[fleqn]{book}
\begin{document}

\newcommand{\av}[1]{\langle #1\rangle}
\newcommand{\Delt}{{\mit\Delta}}

\pagestyle{myheadings}
\markboth{\qquad Asher Peres\hspace*{\fill}}{\hspace*{\fill}
Quaternionic Quantum Interferometry\qquad}

\thispagestyle{empty}
\vspace*{1mm}
{\Large{\bf Quaternionic Quantum Interferometry}}

\bigskip by {\large{\sl Asher Peres}\begin{center}
Department of Physics, Technion---Israel Institute of Technology,\\
32\,000 Haifa, Israel\end{center}}

\begin{quote}{\bf Abstract. } If scattering amplitudes are ordinary
complex numbers (not quaternions) there is a universal algebraic
relationship between the six coherent cross sections of any three
scatterers (taken singly and pairwise). A violation of this relationship
would indicate either that scattering amplitudes are quaternions, or
that the superposition principle fails. Some possible experimental tests
involve neutron interferometry, $K_S$-meson regeneration, and low energy
proton-proton scattering.\end{quote}

When we progress in the hierarchy of numbers, we encounter integers,
real numbers, complex numbers, and then quaternions. The latter are
hypercomplex numbers which can be written as $a+ib+jc+kd$, where
$i^2=j^2=k^2=-1$ and $ij=-ji=k$, etc.  They are the only generalization
of complex numbers that satisfies the associative and distributive laws,
and for which division is possible and unique (Chevalley, 1946). They
were originally introduced in classical physics by Hamilton, in order to
describe 3-dimensional rotations.

When we further progress from classical physics to quantum theory, we
learn that the states of a physical system can be represented by a
linear manifold (Peres, 1993). Namely, if $\psi_1$ and $\psi_2$ are two
possible states of a quantum system, and $c_1$ and $c_2$ are arbitrary
numbers, then the expression $c_1\psi_1+c_2\psi_2$ also represents a
possible state of that system. It is usually taken for granted that the
coefficients $c_1$ and $c_2$ are complex numbers. However, it is
possible to imagine a real quantum theory (Stueckelberg, 1960) or one
based on quaternions (Finkelstein, Jauch, Schiminovich and Speiser,
1962--3; Emch, 1963; Wolff, 1981; Sharma and Coulson, 1987). The purpose
of this article is to show how interferometric experiments can
distinguish between these various quantum theories.

Real quantum theory, although logically consistent, can be easily ruled
out for our world: e.g., complex coefficients are needed in order to
combine linearly polarized photons into circularly polarized ones. More
generally, correspondence with classical physics leads to the
commutation relation $[q,p]=i\hbar$. [Here, it may be pointed out that
Stueckelberg's ``real'' quantum theory requires the introduction of an
operator $J$ satisfying $J^2=-
{\leavevmode\hbox{\small1\kern-3.4pt\normalsize1}}$ and commuting with
all observables. As a consequence, the states $\psi$ and $J\psi$ are
linearly independent, although they are physically indistinguishable.
Moreover, the definition of the scalar product involves $i$ explicitly:
see his equation (A-2.7), page 747.]

A formal test distinguishing between real and complex quantum mechanics
(to be later extended to the case of quaternions) can be performed as
follows. Consider a beam of particles impinging on a scatterer. Let
$\psi_1$ represent the state of the scattered particles, namely,
$\psi_1$ is the difference between the actual state $\psi$ and the state
$\psi_0$ that we would have if the scatterer were absent. Assume that
$\psi_0$ is normalized to unit flux. Now, set a detector at a large
distance $R$ from the scatterer, and let $\chi/R$ represent the state
of a unit flux of particles passing through that detector. Then the
cross section for scattering into that detector is defined as

\[ \sigma_1=|\av{\chi,\psi_1}|^2, \]
where $\av{\chi,\psi}$ denotes the scalar product of the states $\chi$
and $\psi$. If this scalar product is a complex number, we can write

\[ \av{\chi,\psi_1}=a_1\,\exp(i\phi_1), \]
where $a_1$ is real, so that

\[ \sigma_1=a_1{}^2. \]
Similar formulas hold for quaternion quantum theory, with $\exp(i\phi_1)$
replaced by a unimodular quaternion.

Consider now a different scatterer, with scattering amplitude

\[ \av{\chi,\psi_2}=a_2\,\exp(i\phi_2). \]
We have likewise

\[ \sigma_2=|\av{\chi,\psi_2}|^2. \]
Finally, if both scatterers are present, we have to a good approximation

\[ \av{\chi,\psi_{12}}=\av{\chi,\psi_1}+\av{\chi,\psi_2}. \]
This relation is valid if double scattering can be neglected. The total
cross section thus is

\[ \sigma_{12}=|a_1\,\exp(i\phi_1)+a_2\,\exp(i\phi_2)|^2=
 \sigma_1+\sigma_2+2\sqrt{\sigma_1\sigma_2} \cos(\phi_1-\phi_2). \]
Note that $\sigma_{12}$ is well defined provided that the relative
position of the scatterers is held fixed (coherent scattering).

Define

\[ \gamma=(\sigma_{12}-\sigma_1-\sigma_2)/2\,\sqrt{\sigma_1\sigma_2}.\]
This expression involves only observable cross sections, and can
therefore be actually measured for any pair of scatterers. The
measurement of $\gamma$ thus gives a simple criterion for distinguishing
between real and complex quantum theories:

{\sl If $\gamma=\pm1$, real quantum theory is admissible. If
$|\gamma|<1$, we may have complex (or quaternionic) quantum theory. And
if $|\gamma|>1$, the superposition principle is violated.}

Note that the above formulas have been derived for pure states, and they
may not be valid for mixtures, e.g., for an unpolarized beam, if the
cross sections are spin or polarization dependent. For such a mixture,
we can only measure averages:

\[ \av{\sigma_{12}}=\av{\sigma_1}+\av{\sigma_2}+
  2\av{\sqrt{\sigma_1\sigma_2} \cos(\phi_1-\phi_2)}. \]
In that case, we can still define an averaged $\av{\gamma}$ by

\[ \av{\gamma}=(\av{\sigma_{12}}-\av{\sigma_1}-\av{\sigma_2)}
  /2\sqrt{\av{\sigma_1}\,\av{\sigma_2}}
 =\av{\sqrt{\sigma_1\sigma_2} \cos(\phi_1-\phi_2)}
  / {\sqrt{\av{\sigma_1}\,\av{{\sigma_2}}}}. \]
However, this $\av{\gamma}$ is not the cosine of a phase difference and
some of the formulas derived above are not valid. (They do remain valid
if the cross sections are not affected by the spin or polarization
variables).

We now consider a third scatterer and define, as previously, $\sigma_3$,
$\sigma_{31}$, and $\sigma_{32}$, and also

\[ \alpha=(\sigma_{23}-\sigma_2-\sigma_3)/2\,\sqrt{\sigma_2\sigma_3},\]
and

\[ \beta=(\sigma_{31}-\sigma_3-\sigma_1)/2\,\sqrt{\sigma_3\sigma_1}.\]

In complex quantum theory, $\alpha$, $\beta$, and $\gamma$ are the
cosines of $(\phi_2-\phi_3)$, $(\phi_3-\phi_1)$, and $(\phi_1-\phi_2)$,
respectively, and therefore they are not independent, since these angles
sum up to zero. An elementary calculation gives

\[ F(\alpha,\beta,\gamma):=\alpha^2+\beta^2+\gamma^2-
 2\,\alpha\beta\gamma=1. \]
On the other hand, if the amplitudes $\av{\chi,\psi_n}$ are quaternions,
rather than ordinary complex numbers, they do not behave as vectors in a
plane, but as vectors in a four-dimensional space. We then have $0\leq
F(\alpha,\beta,\gamma)\leq1$. The criterion for distinguishing between
complex and quaternionic quantum theory can thus be stated as follows:

{\sl If $F(\alpha,\beta,\gamma)=1$, complex quantum theory is
admissible. If $F<1$, we may have quaternion quantum theory. And if
$F>1$, the superposition principle is violated.}

Note that $F$ can never be negative if $\alpha^2+\beta^2+\gamma^2\leq
3$. It is interesting that no new information can be obtained by
considering the three scatterers simultaneously, because $\sigma_{123}=
\sigma_{12}+\sigma_{23}+\sigma_{31}-\sigma_1-\sigma_2-\sigma_3$ for all
types of quantum theory.

It is thus clear that quaternion quantum theory is essentially different
from complex quantum theory. It is not equivalent to having a hidden
``internal'' degree of freedom (Moravcsik, 1986). Let us examine some
experiments that could distinguish between complex and quaternionic
quantum theories (Peres, 1979).

As explained above, the scatterers must act coherently and multiple
scattering should be negligible. This rules out some tantalizing ideas,
like scattering neutrinos from the three different quarks in baryons.
Conceptually, the simplest test is Bragg scattering by crystals made of
three different kinds of atoms. Indeed this test was performed long ago
with X-rays: the fact that phase angles are coplanar is the basis of the
multiple isomorphous replacement method, used to resolve the structure
of proteins (Blundell and Johnson, 1976). However, X-ray diffraction
involves only the interaction of photons and electrons, and we should
not expect to observe there significant deviations from standard quantum
theory.

On the other hand, nuclear forces are not as well understood as quantum
electrodynamics, and several nontrivial tests can be devised. A simple
one is to examine low energy proton-proton scattering: the Coulomb
amplitude is exactly known, and the nuclear interaction, having short
range, can be analyzed in terms of a few phase shifts (only the $S$-wave
is significant, for low enough energy). If quaternions are implicated in
this process and the $S$-wave phase shift does not involve the same
imaginary unit $i$ as the $i$ that appears in the Coulomb amplitude, it
will be impossible to fit experimental cross sections by the standard
quantum mechanical formulas. Indeed, it was found long ago that there
were ``distressing  irregularities'' in all the data below 10~MeV (Sher,
Signell and Heller, 1970). It is naturally tempting to try to explain
these discrepancies by fitting the data with quaternionic amplitudes.
Unfortunately, this makes the discrepancy even larger! The correct
explanation of this riddle turned out to be a combination of various
relativistic and radiative effects, and corrections for the finite size
of the proton (Peres, 1978).

More promising experiments are those involving only nuclear forces. For
example, a nontrivial test could be neutron diffraction by crystals made
of three different isotopes. Unfortunately, the latter must have large
capture cross sections to give any appreciable ``quaternionic'' effect.
Indeed, the neutron scattering amplitude can be written (Gasiorowicz,
1974) as

\[ f=[\eta\,\sin 2\delta+i\,(1-\eta\,\cos 2\delta)]/2k, \]
where $k$ is the wave number of the neutron, $\delta$ is the $S$-wave
phase shift, and $\eta$ is the elasticity parameter. Both $\delta$ and
$(1-\eta)$ are very small, as can be seen from the formulas for the
scattering and absorption cross sections:

\[ \sigma_s={4\pi\over k^2}\,\left[\eta\,\sin^2\delta+\left(
  {1-\eta\over 2}\right)^2\right], \]
and
\[ \sigma_a=\pi\,(1-\eta^2)/k^2. \]
For thermal neutrons, $4\pi/k^2\simeq 10^8$ b, so that $\delta\simeq
10^{-4}$.

We thus have approximately $f=[\delta+i(1-\eta)/2]/k$, the phase of
which will be nontrivial provided that $(1-\eta)$ has at least the same
order of magnitude as $\delta$. This implies that $\sigma_a$ should be
of the order of $10^4$~b or more, for at least two of the scatterers.
Most materials have much smaller absorption cross sections, and their
$f$ is almost real. As a consequence, we always have
$F(\alpha,\beta,\gamma)\simeq 1$ and this experiment cannot distinguish
between complex and quaternionic quantum theories.

Instead of Bragg scattering, another possibility is neutron
interferometry. The latter involves only the forward-scattering
amplitude, so that this test has less generality, but it is easier to
perform. Consider a plane wave $\exp(i{\bf k\cdot r})$. Passage through
a plate of thickness $L$, having $n$ scatterers per unit volume, changes
the amplitude into $T\,\exp(i\Delt)\exp(i{\bf k\cdot r})$. The
transmission coefficient $T$ is due to reflections at the surfaces of
the plate. The macroscopic phase shift $\Delt$ is given by

\[ \Delt=[\eta\,\sin 2\delta+i\,(1-\eta\,\cos 2\delta)]\,\pi nL/k^2. \]
The real part of $\Delt$ is due to the change in optical path, and the
imaginary part to absorption in the plate. By means of interference with
a reference beam $\exp(i{\bf k'\cdot r})$, with ${\bf k'\simeq k}$, it
is possible to measure both effects.

Now consider two plates made of different materials, taken singly and
jointly. The total transmission coefficient $T_{12}$ will not, in
general, be $T_1T_2$ because of multiple reflections between the plates.
However, the total phase shift $\Delt_{12}$ ought to be
$\Delt_1+\Delt_2$, if our use of complex numbers is legitimate. On the
other hand, quaternion interference usually implies $\Delt_{12}\neq
\Delt_1+\Delt_2$, because quaternion rotations do not commute. We
therefore expect that the interference pattern may be affected by
exchanging the order of two consecutive slabs made of different
materials.

An experimental test, with thick slabs of titanium and aluminum, showed
no such effect (Kaiser, George and Werner, 1984). This is not
surprising, because both metals have low neutron absorption cross
sections, while a strong absorption is needed for seeing quaternionic
effects, as explained above. (The use of highly absorbing materials
would have required a much higher neutron flux.) In spite of its
negative result, this experiment had a remarkable feature: the
introduction of both slabs in the neutron path yielded a higher fringe
visibility that when a single slab was present! The reason for this
curious behavior is that the very large phase shifts that were involved
($+9860^\circ$ for Ti, and $-9980^\circ$ for Al) were an appreciable
fraction of the coherence length of the neutrons. The reduced visibility
of the fringes when only one slab was introduced was due to the partial
lack of coherence of the two neutron paths. That coherence was restored
by the introduction of the second slab, with a nearly opposite phase
shift.

Still another possible test could be a comparison of $K_S$ regeneration
(Perkins, 1972) produced by three different materials, taken singly and
pairwise. Here, the observed quantity is the square of the forward
regeneration amplitude. For our purpose, it is similar to a cross
section, and the expression $F(\alpha,\beta,\gamma)$ can be defined
exactly as before. For this test, it would be especially interesting to
compare neutron-rich and proton-rich nuclei, since they contain
different ratios of up and down quarks. (Yet, if quaternionic effects
occur at the level of individual nucleons, or individual quarks, we
would still have only two different types of scatterers, rather than
three as required.) This experiment has not yet been attempted.

This work was supported by the Gerard Swope Fund, and the Fund for
Encouragement of Research.

\bigskip{\bf References}\frenchspacing

\begin{list}{\null}{\setlength{\itemsep}{0mm}\setlength{\leftmargin}{4mm}}
\item \hspace*{-4mm}Blundell, T. L. and Johnson, L. N. (1976) Protein
crystallography, p.~161. Academic. New York.

\item \hspace*{-4mm}Chevalley, C. (1946) Theory of Lie groups. Princeton
Univ. Press.

\item \hspace*{-4mm}Emch, G. (1963) M\'ecanique quantique
quaternionienne et relativit\'e restreinte. Helv. Phys. Acta
36:739--788.

\item \hspace*{-4mm}Finkelstein, D., Jauch, J. M., Schiminovich S. and
Speiser, D. (1962) Foundations of quaternion quantum mechanics. J. Math.
Phys. 3:207--220.

\item \hspace*{-4mm}Finkelstein, D., Jauch, J. M., Schiminovich S. and
Speiser, D. (1963) Principle of general quaternion covariance. J. Math.
Phys.  4:788--796.

\item \hspace*{-4mm}Gasiorowicz, S. (1974) Quantum physics, p.~384.
Wiley. New York.

\item \hspace*{-4mm}Kaiser, H., George, E. A. and Werner, S. A. (1984)
Neutron interferometric search for quaternions in quantum mechanics.
Phys. Rev A 29:2276--2279.

\item \hspace*{-4mm}Moravcsik, M. J. (1986) Detecting ``secret'' quantum
numbers. Phys. Rev. Lett. 56:908--911.

\item \hspace*{-4mm}Peres, A. (1978) Low energy proton-proton
scattering. Nucl. Phys. A 312:291--296.

\item \hspace*{-4mm}Peres, A. (1979) Proposed test for complex versus
quaternion quantum theory. Phys. Rev. Lett. 42:683--686.

\item \hspace*{-4mm}Peres, A. (1993) Quantum Theory: Concepts and
Methods. Kluwer, Dordrecht.

\item \hspace*{-4mm}Perkins, D. H. (1972) Introduction to high energy
physics, p.~173. Addison-Wesley. Reading.

\item \hspace*{-4mm}Sharma, C. S. and Coulson, T. J. (1987) Spectral
theory for unitary operators on a quaternionic Hilbert space. J. Math.
Phys.  28:1941--1946.

\item \hspace*{-4mm}Sher, M. S., Signell, P. and Heller, L. (1970)
Characteristics of the proton-proton interaction deduced from the data
below 30 MeV. Ann. Physics 58:1--46.

\item \hspace*{-4mm}Stueckelberg, E. C. G. (1960) Quantum theory in real
Hilbert space.  Helv. Phys. Acta 33:727--752.

\item \hspace*{-4mm}Wolff, U. (1981) A quaternionic quantum system.
Phys. Lett.  84A:89--92.

\end{list}
\end{document}